\documentclass{iopart}
\usepackage{graphicx}
\begin{document}

\title[Critical line in undirected Kauffman RBN - the role of percolation]
{Critical line in undirected Kauffman boolean networks - the role
of percolation}

\author{Piotr Fronczak and Agata Fronczak}

\address{Faculty of Physics and Center of Excellence for
Complex Systems Research, Warsaw University of Technology,
Koszykowa 75, PL-00-662 Warsaw, Poland}
\ead{fronczak@if.pw.edu.pl}
\begin{abstract}
We show that to correctly describe the position of the critical
line in the Kauffman random boolean networks one must take into
account percolation phenomena underlying the process of damage
spreading. For this reason, since the issue of percolation
transition is much simpler in random undirected networks, than in
the directed ones, we study the Kauffman model in undirected
networks. We derive the mean field formula for the critical line
in the giant components of these networks, and show that the
critical line characterizing the whole network results from the
fact that the ordered behavior of small clusters shields the
chaotic behavior of the giant component. We also show a possible
attitude towards the analytical description of the shielding
effect. The theoretical derivations given in this paper quite
tally with numerical simulations done for classical random graphs.
\end{abstract}


\section{Introduction}

Almost 40 years ago Stuart Kauffman proposed random Boolean
networks (RBNs) for modelling gene regulatory networks
\cite{kauffman1969}. Since then, beside its original purpose, the
model and its modifications have been applied to many different
phenomena like cell differentiation \cite{huang2000}, immune
response \cite{kauffman1989}, evolution \cite{bornholdt1998}, opinion formation \cite{lambiotte},
neural networks \cite{wang1990}, and even quantum gravity problems
\cite{baillie1994}.

The original RBNs were represented by a set of $N$ elements,
$\sum_t=\{\sigma_1(t), \sigma_2(t),...,\sigma_N(t)\}$, each element
$\sigma_i$ having two possible states: active ($1$), or inactive ($0$). The
value of $\sigma_i$ was controlled by $k$ other elements of the
network, i.e.
\begin{equation}\label{fi}
\sigma_i(t+1)=f_i(\sigma_{i_1}(t),\sigma_{i_2}(t),...,\sigma_{i_k}(t)),
\end{equation}
where $k$ was a fixed parameter.
The functions $f_i$ were selected so that they have returned
values $1$ and $0$ with probabilities respectively equal to $p$
and $1-p$. The parameters $k$ and $p$ have determined the dynamics
of the system (Kauffman network), and it has been shown that for a
given probability $p$, there exists the critical number of inputs
\cite{derrida1986}
\begin{equation}\label{eq_kauffman}
k_c=\frac{1}{2p(1-p)},
\end{equation}
below which all perturbations in the initial state of the system
die out ({\it frozen phase}), and above which a small perturbation
in the initial state of the system may propagate across the entire
network ({\it chaotic phase}).

In fact, the behavior of Kauffman model in the vicinity of the
critical line $k_c(p)$ has become a major concern of scientists
interested in gene regulatory networks. The main reason for this
was the conjecture that living organisms operate in a region
between order and complete randomness or chaos (the so-called {\it
edge of chaos}) where both complexity and rate of evolution are
maximized \cite{kauffman1990,sole2001,stauffer1994}. The analogous
behavior has been noticed in Kauffman networks, which in the
interesting region described by eq. (\ref{eq_kauffman}) show
stability, homeostatis, and the ability to cope with minor
modifications when mutated. The networks are stable as well as
flexible in this region.

Recently, when data from real networks have become available
\cite{albert2002,barabasi2002}, a quantitative comparison of the
{\it edge of chaos} in these datasets and RBN models has brought
an encouraging and promising message that even such simple model
may quite well mimic characteristics of real systems.

Since, however, one has noticed that real genetic networks exhibit
a wide range of connectivities, the recent modifications of the
standard RBN take into consideration a distribution of nodes'
degrees $P(k)$. It has been shown that if the random topology of
the directed network is homogeneous (i.e. all elements of the
network are statistically equivalent), then the network topology
can be meaningfully characterized by the average in-degree
$\langle k\rangle$, and the transition between frozen and chaotic
phase occurs for \cite{sole1997}:
\begin{equation}\label{eq_sole}
\langle k\rangle_c=\frac{1}{2p(1-p)}.
\end{equation}

Several authors \cite{lee,boguna} have provided a
general formula for the edge of chaos in directed networks
characterized by the joint degree distribution $P(k,q)$
\begin{equation}\label{lee_eq}
\frac{\langle kq\rangle}{\langle q\rangle}=\frac{1}{2p(1-p)},
\end{equation}
where $k$ and $q$ correspond to in- and out-degrees of the same
node, respectively. The formula (\ref{lee_eq}) shows that the
position of the critical line depends on the correlations between
$k$ and $q$ in such networks. It is also easy to show that the
previous results (\ref{eq_kauffman}) and (\ref{eq_sole})
immediately follow from (\ref{lee_eq}) if one assumes the lack of
correlations $P(k,q)=P_{in}(k)P_{out}(q)$.

Very recently, it has been shown by finite size scaling methods
(FSS) that the critical connectivity $\langle k\rangle_c^{FSS}$
significantly deviates from the value established by the
Eq.~(\ref{eq_sole}), even for large system sizes \cite{rohlf2007}.
More precisely, one observes that $\langle
k\rangle_c^{FSS}<\langle k\rangle_c$. To support the observation
the authors recall other studies \cite{ramo2006} which suggest
that gene regulatory networks appear to be in the ordered regime
and reside slightly below the phase transition between order and
chaos in opposite to the theory which proposes the critical line
to be an evolutionary attractor.

In the present paper, we suggest other explanation of the observed
discrepancy. We show (both analytically and numerically) that the
discrepancies are due to the percolation phenomena, which become
important in the region of small values of the parameter $\langle
k\rangle$.

\begin{figure}[h]

\begin{center}
\includegraphics[width=5cm]{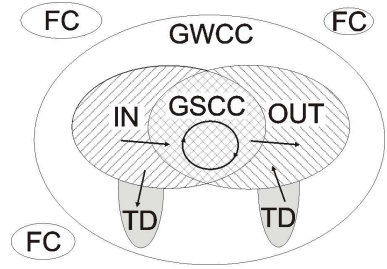}
\end{center}
\caption{General structure of a directed network above the percolation threshold.} \label{fig1}
\end{figure}

To understand the complexity of percolation phenomena in directed
graphs let us recall the structure of such a graph
\cite{newman2001,dorogov2001}. In general, a directed graph
consists of a giant weakly connected component (GWCC) and several
finite components (FCs). In the GWCC every site is reachable from
every other, provided that the links are treated as bidirectional.
The GWCC is further divided into a giant strongly connected
component (GSCC), consisting of all sites reachable from each
other following directed links. All sites reachable from the GSCC
are referred to as the giant OUT component, and the sites from
which the GSCC is reachable are referred to as the giant IN
component. The GSCC is the intersection of the IN and OUT
components. All sites in the GWCC, but not in the IN and OUT
components, are referred to as the tendrils (TDs) (see Fig. \ref{fig1}).

\begin{figure}[h]
\begin{center}
\includegraphics[width=13cm]{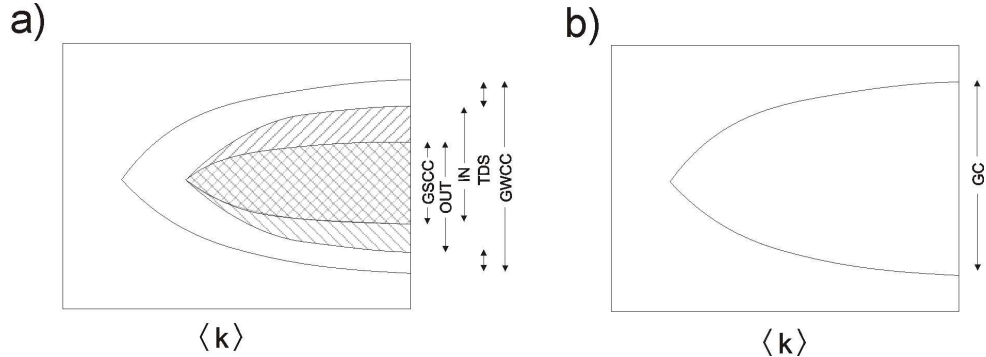}
\end{center}
\caption{Schematic plot of sizes of network components as a
function of average node degree in a) directed ER graphs and b)
undirected ER graphs.}  \label{fig2}
\end{figure}

Size of all components listed above has doubtless impact on
propagation of perturbations in directed RBNs. Moreover, GSCC and
GWCC start to form at different values of the parameter $\langle
k\rangle$ (see Fig. \ref{fig2}a). Although it has been shown
\cite{newman2001,dorogov2001} how to find the relative sizes of
the components (for example GWCC appears when $\langle kq\rangle
\geq \langle q\rangle$), the problem of how to implement the
results to the theory of perturbation spreading in RBNs is still
far from being solved. To make the first step in this direction,
and to show the importance of percolation phenomena on dynamics of
RBN we concentrate on undirected case of the model. Although the
original RBNs have been defined as directed ones, the study of
undirected networks significantly reduces complexity of the
problem (see Fig.~\ref{fig2}).

To this end, we organize the paper as follows. In the next
section, we present numerical methodology and finite-size scaling
of perturbation spreading in RBNs. In section 3 we derive general
relation describing position of the critical line in undirected
RBNs with arbitrary distribution of connections $P(k)$, in the
analogy of the mean-field theory for directed RBNs
\cite{derrida1986}. Comparing the theory with numerical
simulations we show significant deviations between the both
approaches. Then an improved treatment including percolation
phenomena is presented in section 4. A summary of our findings is
given in section 5.

\section{Critical line in undirected random graphs - numerical simulations}

In order to find the position of the critical line in RBN one has
to examine the sensitivity of its dynamics with regard to initial
conditions. In numerical studies such a sensitivity can be
analyzed quite simply. One has to start with two initial states
$\sum_0=\{\sigma_1(0), \sigma_2(0),...,\sigma_N(0)\}$ and
$\widetilde{\sum}_0=\{\widetilde{\sigma}_1(0),
\widetilde{\sigma}_2(0),...,\widetilde{\sigma}_N(0)\}$, which are
identical except for a small number of elements, and observe how
the differences between both configurations $\sum_t$ and
$\widetilde{\sum}_t$ change in time. If a system is robust then
the studied configurations lead to similar long-time behavior,
otherwise the differences develop in time. A suitable measure for
the distance between the configurations is the overlap $x(t)$
defined as
\begin{equation}\label{saf1}
x(t)=1-\frac{1}{N}\sum_{i=1}^N|\sigma_i(t)-\widetilde{\sigma}_i(t)|.
\end{equation}
Note, that in the limit $N\rightarrow\infty$, the overlap becomes
the probability for two arbitrary but corresponding elements,
$\sigma_i(t)$ and $\widetilde{\sigma_i}(t)$, to be equal.
Moreover, the stationary long-time limit of the overlap
$x=\lim_{t\rightarrow\infty}x(t)$ can be treated as the order
parameter of the system. If $x=1$ then the system is insensitive
to initial perturbations (frozen phase), while for $x<1$, the
initial perturbations propagate across the entire network (chaotic
phase).

\begin{figure}[h]
\begin{center}
\includegraphics[width=11cm]{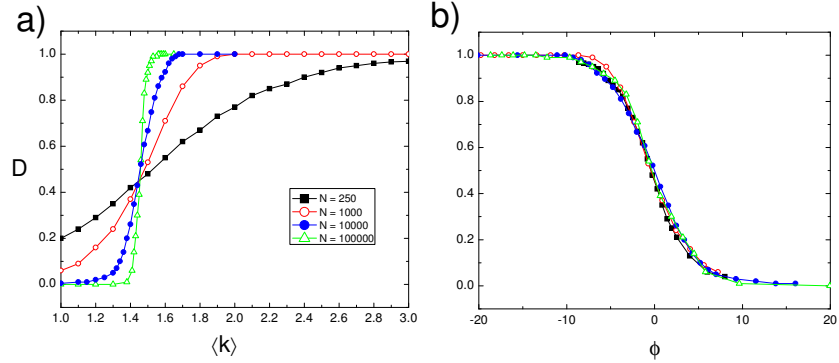}
\end{center}
\caption{Probability $D$ against a) control parameter $\langle k
\rangle$ and b) rescaled parameter $\phi$ for $p=0.5$.} \label{fig3}
\end{figure}

For numerical purposes we define the probability $D$ that the
system is sensitive to perturbations
\begin{equation}
D=\frac{\sum_{x(t=T)<x(0)}^R 1}{R},
\end{equation}
where $R$ is the number of generated networks, and $T$ is the
number of system updates. In our simulations we take $RN=10^6$ and
$T=200$. The Fig. \ref{fig3}a presents a typical example of $D$
dependence on our control parameter $\langle k \rangle$ for
different network sizes. Then, we apply finite-size scaling method
\cite{barber1983} to determine how the probability $D$ scales with
the system size. Around some critical point, we predict that
systems of all sizes are indistinguishable except for a change of
scale. This suggests
\begin{equation}
\label{fss}
D(\langle k \rangle)=f(\phi),
\end{equation}
where
\begin{equation}
\phi=\left( \frac{\langle k \rangle-\langle k \rangle_c}{\langle k
\rangle_c}\right) N^{1/\nu}.
\end{equation}
In Eq.~(\ref{fss}), $f$ is one of the functions shown in the
figure 3a, $\langle k \rangle_c$ is the critical point, and
$N^{1/\nu}$ provides the change of scale. Fig. \ref{fig3}b shows
how the probability $D$ depends on the parameter $\phi$ with
fitted parameters $\langle k \rangle_c=1.45\pm 0.04$ and
$\nu=2.2\pm 0.1$.

\begin{figure}[h]
\begin{center}
\includegraphics[width=6cm]{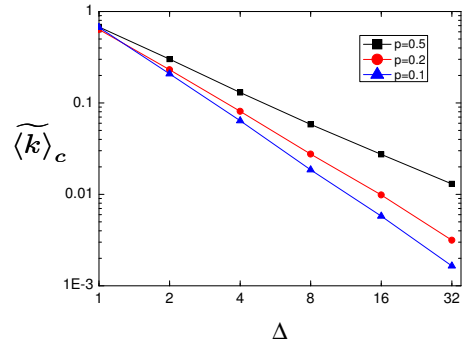}
\end{center}
\caption{Normalized critical connectivity against the number of
perturbed nodes in networks of $N=1000$ elements. Lines are shown
only for better visibility of the presented dependence.} \label{fig4}
\end{figure}

The other problem which should be noted here is the observation
that $\langle k \rangle_c$ depends on the number of initially
perturbed nodes. In the Fig. \ref{fig4} we plot the dependence of
normalized critical connectivity
\begin{equation}
\widetilde{\langle k \rangle}_c=\frac{\langle k \rangle_c(\Delta)-\langle k \rangle_c}{\langle k \rangle_c},
\end{equation}
against the number $\Delta$ of initially perturbed nodes in the
network of $N=1000$ elements. For further calculations we choose
$\Delta=0.032\;N$, since then the error in $\langle k \rangle_c$
is less than the error arising in finite-size scaling.

\begin{figure}[h]
\begin{center}
\includegraphics[width=6cm]{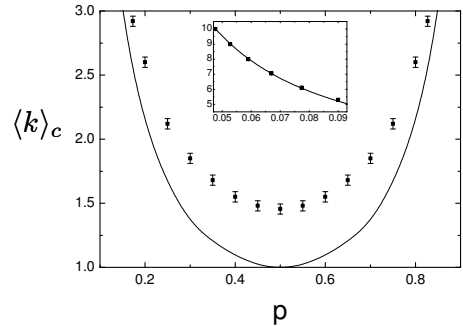}
\end{center}
\caption{Phase diagram for undirected RBN model. Points show
results obtained by numerical simulations. The line is a solution of
eq. (\ref{af10}).} \label{fig5}
\end{figure}

In Fig.~\ref{fig5}, using the method described above, we show the
numerically obtained values of $\langle k \rangle_c$ against the
parameter $p$. For $p=0.5$ critical connectivity is minimal, i.e.
$\langle k \rangle_c=1.45$. Please note that the size of the giant
component for this connectivity is about one half of the whole
network. One can expect that a large number of isolated nodes and
clusters can significantly affect the perturbation spreading rate
in this regime. Moreover, it has been demonstrated \cite{bialas},
that the giant component is correlated in sparse networks. In the
following, we will show that a mean field theory which does not
take into account these percolation and correlation issues,
although correct for large values of $\langle k \rangle$, deviates
from numerical results for $\langle k \rangle$ close to $1$.

\section{Damage spreading in undirected Kauffman RBN - a simple approach}

In this section, we derive a mean field formula for the critical
line characterizing Kauffman boolean model in undirected and
uncorrelated random graphs with arbitrary degree distributions
$P(k)$. To this end, we partially reproduce and generalize a
simple annealed calculations that have been for the first time
carried out by Derrida and Pomeau \cite{derrida1986}. The case of
random directed networks has been studied by Aldana
\cite{aldana2003physd}, and also by Lee and Rieger \cite{lee}.

Thus, let $x_i(k,t)$ corresponds to the probability that a given
element $i$ of degree $k$ possesses the same value in both
configurations $\sum_t$ and $\widetilde{\sum}_t$ of the considered
boolean network, i.e. $\sigma_i(t)=\widetilde{\sigma}_i(t)$. It
occurs either when all the $k$ inputs of $\sigma_i(t)$ are equal
to respective inputs of $\widetilde{\sigma}_i(t)$, or when the
function $f_i$, cf. (\ref{fi}), ascribed to the node $i$ returns
the same value for these two configurations. The first case
happens with probability
\begin{equation}\label{af0}
X(q_1,q_2,\dots,q_k,t-1)=x(q_1,t-1)x(q_2,t-1)\dots x(q_k,t-1),
\end{equation} where
$x(q_j,t-1)$ represents probability that in the previous time step
$(t-1)$ the $j$th nearest neighbor of $i$ having degree $q_j$ was
in the same state in the two considered configurations. It is also
easy to see that the second case arises with probability
$p^2+(1-p)^2$, when at least one of the $k$ inputs of $\sigma_i$
differs from its counterpart in $\widetilde{\sigma}_i$ giving rise
to the same values of $\sigma_i$ and $\widetilde{\sigma}_i$. Such
a situation, in turn, happens with probability equal to
$1-X(q_1,q_2,\dots,q_k,t-1)$. Taking all the above together we
find that the probability $x_i(k,t)$ that
$\sigma_i(t)=\widetilde{\sigma}_i(t)$ is given by
\begin{eqnarray}\label{af1}
x_i(k,t+1)&=&X(q_1,\dots,t)+(p^2+(1-p)^2)X(q_1,\dots,t)\nonumber
\\&=&1-2p(1-p)\left(1-X(q_1,q_2,\dots,q_k,t)\right),
\end{eqnarray}
where $q_1,q_2,\dots,q_k$ stand for degrees of nodes found in the
nearest neighborhood of the node $i$.

The equation (\ref{af1}) describes dynamics of a single node $i$
of degree $k$. In order to study boolean dynamics of the whole
network one has to average the equation, first over the nearest
neighborhood of $i$, next over the whole network. The first step
simply means averaging over the distribution
$P(q_1,q_2,\dots,q_k/k)$, which describes probability that nearest
neighbors of $i$ have degrees respectively equal to
$q_1,q_2,\dots,q_k$
\begin{equation}\label{af2}
x(k,t+1)=1-2p(1-p)\left(1-\sum_{q_1,\dots,q_k}X(q_1,\dots,
t)P(q_1,\dots/k)\right),
\end{equation}
whereas the second step corresponds to averaging of the last
equation over the node degree distribution $P(k)$ characterizing
the whole network. Note, that we have omitted the subscript $i$ at
$x(k,t+1)$ in Eq.~(\ref{af2}). After averaging, $x(k,t+1)$ refers
to the set of nodes having the same degree $k$.

At the moment, before we proceed with our calculations let us
outline structural properties of the studied networks. At the
beginning let us remind that the assumed lack of higher-order
correlations (e.g. three-point or four-point correlations) means
that a given link $\{i,j\}$ does not influence other links of the
considered nodes $i$ and $j$. It translates to the fact that the
conditional probability $P(q_1,q_2,\dots,q_k/k)$ factorizes
\begin{equation}\label{af4}
P(q_1,q_2,\dots,q_k/k)=P(q_1/k)P(q_2/k)\dots P(q_k/k),
\end{equation}
where $P(q_j/k)$ describes probability that a node of degree $q_j$
is the nearest neighbor of a node having degree $k$. Given the
formulas (\ref{af0}), (\ref{af4}) and (\ref{af5}), the equation
(\ref{af2}) further simplifies as follows
\begin{equation}\label{af6}
x(k)=1-2p(1-p)\left(1-\left(\sum_qx(q)P(q/k)\right)^k\right),
\end{equation}
where, since we are interested in the stationary (i.e. for
$t\rightarrow\infty$) solutions of this equation, we have omitted
dependence on time $t$.

\begin{figure}[h]
\begin{center}
\includegraphics[width=6cm]{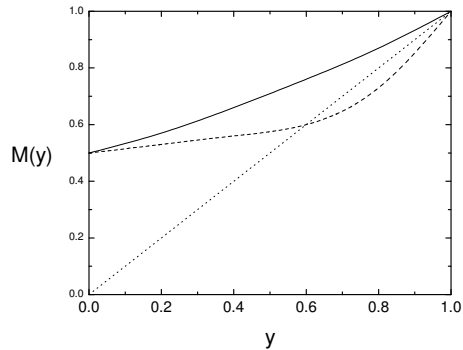}
\end{center}
\caption{The map $y=M(y)$ considered in the text. The solid line
corresponds to the situation when the only stable solution is
$\langle kx\rangle = \langle k\rangle$, i.e. $x(k)=1$ for all
values of $k$. The dashed line shows the case when the second
solution $\langle kx\rangle < \langle k\rangle$ appears.}
\label{fig6}
\end{figure}

Now, assuming the lack of two point correlations, i.e.
\begin{equation}\label{af5}
P(q_j/k)=\frac{q_j}{\langle k\rangle}P(q_j),
\end{equation}
which causes that the nearest neighborhood of each node is the
same (in statistical terms), and then multiplying both sides of
Eq.~(\ref{af6}) by $k$, and finally averaging the resulting
equation over the node degree distribution $P(k)$, we get the
desired mean-field equation which describes stationary states of
the Kauffman model defined on undirected and uncorrelated random
networks with arbitrary degree distributions
\begin{equation}\label{af7}
\frac{\langle kx\rangle}{\langle k\rangle}=M\left(\frac{\langle
kx\rangle}{\langle k\rangle}\right)=1-2p(1-p)\left(1-\sum_k
\left(\frac{\langle kx\rangle}{\langle
k\rangle}\right)^k\frac{k}{\langle k\rangle}P(k)\right),
\end{equation}
where $\langle kx\rangle=\sum_kkx(k)P(k)$.

At the moment, note that the state $\langle kx\rangle=\langle
k\rangle$, which in fact corresponds to the set of conditions
$x(k)=1$ for all nodes' degrees $k$, is always a solution of the
last equation, see Fig. \ref{fig6}. Note also, that this solution may be
stable or unstable depending on properties of the considered map
$y=M(y)$, where $y=\langle kx\rangle/\langle k\rangle$
(\ref{af7}). In fact, one can show that the solution loses its
stability, when another solution $\langle kx\rangle<\langle
k\rangle$ of this equation appears. For the first time it happens
when
\begin{equation}\label{af8}
\lim_{y\rightarrow 1^-} \frac{dM(y)}{dy}=1,
\end{equation}
where the limit $y\rightarrow 1^-$ is equivalent to $\langle
kx\rangle\rightarrow\langle k\rangle^{-}$. Substituting
(\ref{af7}) into (\ref{af8}) we get the condition for the phase
transition between ordered and chaotic behavior of the Kauffman
model defined on undirected and uncorrelated random network
\begin{equation}\label{af9}
\frac{\langle k^2\rangle}{\langle k\rangle}=\frac{1}{2p(1-p)},
\end{equation}
where $\langle k\rangle$ and $\langle k^2\rangle$ stand for the
first and the second moment of the degree distribution $P(k)$,
respectively. In the following we briefly analyze the formula for
the critical line (\ref{af9}) in classical random graphs. The case
of scale-free networks $P(k)\sim k^{-\gamma}$, for which the
second moment $\langle k^2 \rangle$ of the degree distribution
becomes important, has been analyzed in \cite{fronczak2007}.

Thus, since in classical random graphs $\langle k^2
\rangle=\langle k \rangle^2+\langle k \rangle$, the formula
(\ref{af9}) simplifies
\begin{equation}\label{af10}
\langle k\rangle_c=\frac{1}{2p(1-p)}-1.
\end{equation}
In the Fig. \ref{fig5} one can see numerical simulations of the
Kauffman boolean model defined on these graphs as compared with
the expression (\ref{af10}). In our previous paper \cite{fronczak2007} we have
suggested that the visible discrepancy between numerical
calculations and their theoretical prediction for $\langle k
\rangle \rightarrow 1$ (i.e. for $p\rightarrow 0.5$) may result
from the fact that $\langle k \rangle =1$ corresponds to the
percolation threshold in these networks. A simple heuristic
argument behind this statement was the following: because the size
of the largest component near $\langle k \rangle =1$ is
significantly smaller than the network size (the network is
divided into several disconnected components), any perturbation
cannot propagate across the entire system, and the frozen phase is
easier achieved. It means that the closer percolation threshold
$\langle k\rangle=1$ we are, the more crumbled network (separated
pieces of the whole system) we analyze, and the theoretical
prediction given by Eq.~(\ref{af10}) works worse and worse. In
fact, comparing the general formula $\langle k^2\rangle/\langle
k\rangle=2$ \cite{molloy1995} for the percolation threshold in arbitrary
undirected and uncorrelated random network with the general
expression for the critical line (\ref{af9}), one can show that
the arguments exposed in relation to classical random graphs
should also apply for the whole class of the considered networks.

In the next section we show how to adjust the approach presented
in this section in order to correctly describe properties of the
analyzed systems in the whole range of parameters, also in the
vicinity of the percolation transition.

\section{The effect of percolation phenomena on damage spreading}

In the following, in order to correctly address the problem of
damage spreading in the vicinity of percolation transition, that
has been outlined at the end of the previous section, we use a few
important results on percolation phenomena in the considered class
of networks. To begin with, we recall these results. As we are
going to directly (i.e. in the course of numerical simulations)
check our derivations in classical random graphs, together with
general formulas describing behavior of arbitrary undirected and
uncorrelated random networks we also provide the respective
formulas for these graphs.

Thus, as we have already mentioned, random graph with a given node
degree distribution $P(k)$ does not need to be connected. However,
if
\begin{equation}\label{af11}
\frac{\langle k^2\rangle}{\langle k\rangle}>2,
\end{equation}
that in classical random graphs translates into
\begin{equation}\label{af12}
\langle k\rangle>1,
\end{equation}
the giant component $GC$ emerges which gathers a finite fraction
of all nodes and links. The size of the giant component $S$, i.e.
the probability that an arbitrary node belongs to $GC$, is given
by the below formula
\begin{equation}\label{af13}
S=1-G_0(u),
\end{equation}
where $u$ is the solution of the self-consistency equation
\begin{equation}\label{af14}
u=G_1(u),
\end{equation}
and $1-u^2$ is the probability that a link belongs to the giant
component. The functions $G_0(u)$ and $G_1(u)$ correspond to
generating functions of the node degree distribution $P(k)$, and
the conditional distribution $P(q_j/k)$ (\ref{af5}), respectively.
Since in classical random graphs $G_0(x)=G_1(x)=e^{\langle
k\rangle(x-1)}$, the formula (\ref{af13}) for these networks
significantly simplifies
\begin{equation}\label{af15}
S=1-e^{-\langle k\rangle S},
\end{equation}
and the expression for $u$ becomes
\begin{equation}\label{af16}
u=1-S.
\end{equation}

The general results on percolation transition in random undirected
and uncorrelated networks outlined in the previous paragraph are
already well-known. They have been derived by several authors
using different theoretical approaches, see e.g. \cite{molloy1995, cohen2000}. Recently,
however, a new interesting results completing our knowledge in
this subject have been obtained by Bia\l as and Ole\'s \cite{bialas}. The
authors have shown that the neighboring nodes in the giant
connected components are disassortatively correlated. They have
also derived analytic formulas for the node degree distribution
\begin{equation}\label{af17}
P^{*}(k)=P(k)\frac{1-u^k}{S},
\end{equation}
and the joint nearest-neighbor degree distribution
\begin{equation}\label{af18}
P^{*}(k,q)=P(k,q)\left(\frac{1-u^{k+q-2}}{1-u^2}\right)=\frac{kP(k)qP(q)}{\langle
k\rangle^2}\left(\frac{1-u^{k+q-2}}{1-u^2}\right),
\end{equation}
characterizing the giant component. Let us note, that in the limit
$u\rightarrow 0$, when the giant component covers the whole
network $S\rightarrow 1$, the both distributions $P^{*}(k)$ and
$P^{*}(k,q)$ respectively converge to distributions $P(k)$ and
$P(k,q)$, which characterize random uncorrelated networks. The
formulas (\ref{af17}) and (\ref{af18}) are crucial for the further
developments of this paper, as they show that although in average
the considered networks are uncorrelated, in the vicinity of
percolation transition their giant components are disassortative
(note that we still do not know anything about higher-order
correlations in $GC$s). Now, since we know that this type of
correlations makes different spreading-like phenomena more
difficult \cite{newman2002}, we expect that disassortativity of the giant
component is partially responsible for the discrepancy observed in
Fig. \ref{fig5}, with the crumbling of the system as a whole being the
second reason. Below, we show that taking these effects into
consideration significantly improves theoretical prediction for
the critical line in the Kauffman model defined on random
uncorrelated networks.

Thus, let us study damage spreading within the giant component of
the considered networks. Knowing properties of this cluster, we
can start our analysis from Eq.~(\ref{af6}), which is valid for
the general class of networks with two-point correlations. The
conditional probability $P^{*}(q/k)$ for the giant component can
be calculated from the standard expression \cite{boguna2003}
\begin{equation}\label{af19}
P^{*}(q/k)=\frac{\langle k\rangle^{*}P^{*}(k,q)}{kP^{*}(k)},
\end{equation}
where
\begin{equation}\label{af20}
\langle k\rangle^{*} =\sum_kkP^{*}(k)=\langle
k\rangle\frac{1-u^2}{S},
\end{equation}
is the average degree characterizing this component. Inserting
(\ref{af17}) and (\ref{af18}) into (\ref{af19}) we get
\begin{equation}\label{af21}
P^{*}(q/k)=P(q/k)\left(\frac{1-u^{k+q-2}}{1-u^k}\right),
\end{equation}
where $P(q/k)$ is given by (\ref{af5}). The last formula
(\ref{af21}) can be also written in the equivalent form
\begin{equation}\label{af22}
P^{*}(q/k)=P^{*}(q)\left(\frac{q}{\langle
k\rangle}\frac{S}{(1-u^q)}\right)\left(\frac{1-u^{k+q-2}}{1-u^k}\right),
\end{equation}
which turns out to be useful in our further developments.

Now, let us apply the equation (\ref{af6}) to the giant component
\begin{equation}\label{af23}
x^*(k)=1-2p(1-p)\left(1-\left(\sum_qx^*(q)P^*(q/k)\right)^k\right).
\end{equation}
Due to the complicated form of the conditional distribution
$P^*(q/k)$ (\ref{af21}), it is impossible to deduce on possible
solutions of the equation (\ref{af23}) in the same way as we have
done it for the case of uncorrelated networks. However,
substituting (\ref{af22}) into (\ref{af23}) we obtain
\begin{equation}\label{af24}
x^*(k)=1-2p(1-p)\left(1-\left(\sum_q\kappa(q)w(q,k)P^*(q)\right)^k\right),
\end{equation}
where
\begin{equation}\label{af25}
\kappa(q)=x^*(q)\frac{S}{1-u^q}\frac{q} {\langle k\rangle},
\end{equation}
and
\begin{equation}\label{af26}
w(q,k)=\frac{1-u^{q+k-2}}{1-u^k}.
\end{equation}
Next, applying a mean field approximation to Eq.~(\ref{af24})
\begin{eqnarray}\label{af27}
\langle\kappa(q)w(q,k)\rangle^*&=&\sum_q\kappa(q)w(q,k)P^*(q)\\
&\simeq&\left(\sum_q\kappa(q)P^*(q)\right)\left(\sum_qw(q,k)P^*(q)\right)
=\kappa^*w^*(k)\nonumber,
\end{eqnarray}
we get the simplified equation
\begin{equation}\label{af28}
x^*(k)=1-2p(1-p)\left(1-\left(\kappa^*w^*(k)\right)^k\right),
\end{equation}
which after some algebra, consisting in multiplying both sides of
this equation by $(k/\langle k\rangle)(S/(1-u^k))$ and then
averaging it over $P^*(k)$, further simplifies and becomes
equivalent to Eq.~(\ref{af7})
\begin{equation}\label{af29}
\kappa^*=M^*(\kappa^*)=1-2p(1-p)\left(1-\sum_k\left(\kappa^*w^*(k)\right)^k\frac{k}{\langle
k\rangle}P(k)\right).
\end{equation}

The equivalence of the two equations (\ref{af7}) and (\ref{af29})
is visible when $u\rightarrow 0$ (i.e. $S\rightarrow 1$). Then,
the parameter $\kappa^*$, see Eqs.~(\ref{af25}) and (\ref{af27}),
simplifies as follows
\begin{eqnarray}
\kappa^*&=&\sum_{k}x^*(k)\frac{k}{\langle
k\rangle^*}\frac{S}{(1-u^q)}P^*(k)\label{af29a}\\&\simeq&\sum_{k}x^*(k)\frac{k}{\langle
k\rangle^*}P^*(k)\simeq\frac{\langle xk\rangle}{\langle
k\rangle}\label{af29b},
\end{eqnarray}
where the averages $\langle\dots\rangle^*$ and
$\langle\dots\rangle$ have their standard meaning (in our
calculations $'*'$ always refers to the giant component). This
equivalence, also makes possible a similar analytical treatment of
Eq.~(\ref{af29}), as the one performed in the reference case of
uncorrelated networks, compare Eqs.~(\ref{af6})-(\ref{af9})).

Thus, in order to find condition for the transition between
ordered and chaotic phase of the Kauffman model defined in giant
components of random uncorrelated networks we have to check when
the solution $\kappa^*=1$ (\ref{af29a}), corresponding to
$x^*(k)=1$ for all nodes' degrees, becomes unstable. In fact, it
happens when
\begin{equation}\label{af30}
\lim_{\kappa^*\rightarrow 1^-}\frac{dM^*(\kappa^*)}{d\kappa^*}=1.
\end{equation}
From the equation (\ref{af29}) it follows that the condition has a
very simple form
\begin{equation}\label{af31}
\frac{\langle k^2w^*(k)^k\rangle}{\langle
k\rangle}=\frac{1}{2p(1-p)},
\end{equation}
where
\begin{equation}\label{af31a}
w^*(k)=\sum_{q}w(q,k)P^*(q)=\sum_{q}\frac{1-u^{q+k-2}}{1-u^k}P^*(q)
\end{equation}
is defined in Eq.~(\ref{af27}). At the moment, let us note that in
the limiting case of $u\rightarrow 0$, the parameter
$w^*(k)\rightarrow 1$, and the formula (\ref{af31}) simplifies to
the previous condition (\ref{af9}).

\begin{figure}[h]
\begin{center}
\includegraphics[width=8cm]{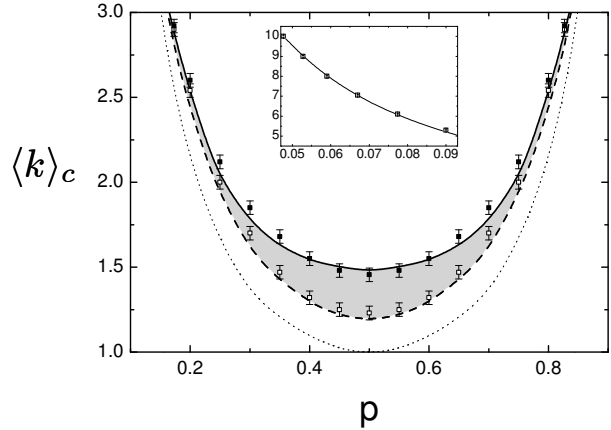}
\end{center}
\caption{Phase diagram for undirected RBN model in classical
random graphs. Dotted line is a solution of basic
Eq.~(\ref{af10}). Filled points represent numerical simulations
made for the whole network (the same data are shown in Fig. 5).
Open points and dashed line correspond respectively to numerical
simulations and analytic prediction of Eq.~(\ref{af31}) for the
Kauffman model defined in giant components only. Solid line is the
solution of final eq. (\ref{final}). Gray area emphasizes the set
of parameters where the chaotic behavior, although present in the
giant component, is not yet visible in the whole network.}
\label{fig7}
\end{figure}

It is easy to check, that in the simplest case of classical random
graphs the parameter $w^*(k)$ (\ref{af31a}) is given by
\begin{equation}\label{af32}
w^*(k)=\frac{1-u-u^{k-1}+u^{k+u-1}}{(1-u)(1-u^k)}.
\end{equation}
Inserting the last formula into (\ref{af31}), and then numerically
solving the resulting equation for $\langle k\rangle$ we obtain
theoretical prediction for the critical line of the Kauffman model
in giant components of these graphs. In Fig. \ref{fig7} one can see that
numerical simulations quite tally with the theoretical prediction
of Eq.~(\ref{af31}). Given the figure, we would also like to take
note of two other interesting effects related to the analyzed
problem. First, the critical line characterizing the giant
component significantly differs from the curve described by the
formula (\ref{af9}). It is shifted towards the numerically
obtained critical line characterizing the whole network. The
observation is in some sense promising, as it partially confirms
the main proposition of this paper, which states that the
percolation transition is responsible for discrepancies observed
in Fig. \ref{fig5}. The second effect concerns mutual relationship between
the behavior of the giant component and the behavior of the whole
network. Since one knows that the giant component makes up a
macroscopic part of the network (it grows linearly with the
network size $N$, and becomes infinite in the thermodynamic limit
$N\rightarrow\infty$) one could expect that dynamics of the whole
network should reflect behavior of the giant component. Thus, the
question is, why the numerically obtained critical line
characterizing the whole network differs from the theoretical
prediction for the giant component. In other words, why, for the
set of parameters marked by the light gray area in Fig. \ref{fig7}, the
chaotic behavior of the giant component is not visible in the
whole network.

To solve the problem stated at the end of the last paragraph, let
us briefly recall what the numerical simulations of the Kauffman
model consist in. Thus, in numerical studies we check
how the initial perturbation of the system $x(0)\equiv 1-\Delta$
(\ref{saf1}), where $\Delta\ll 1$, develops over time. In general,
when the parameter $x(t=T)<x(0)$ we identify the system as the
chaotic one. On the other hand, when $x(0)\leq
x(t=T)\leq 1$ we treat it as being in the ordered phase. In reality,
however, due to the fact that in the vicinity of the percolation
transition the considered Kauffman networks are strongly
heterogenous, they consist of the giant component which is
escorted by a number of small tree-like clusters and isolated
nodes, the systems should by treated more carefully.

\begin{figure}[h]
\begin{center}
\includegraphics[width=8cm]{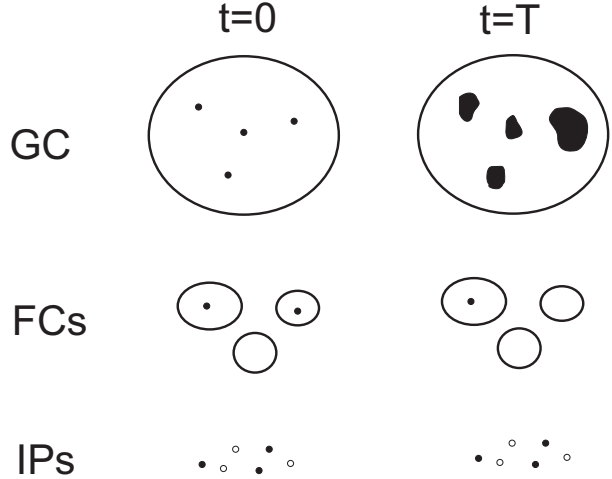}
\end{center}
\caption{Schematic plot of spreading of perturbation in giant component (GC), finite clusters (FCs) and isolated points (IPs). In GC damage spreads, in FCs it shrinks, while in IPs it does not change.}  \label{fig8}
\end{figure}

To better describe the situation let us choose the system
parameters from the region that is marked by the light gray color in Fig. \ref{fig7}. Then, we introduce a quantity $\Omega$, which measures {\it chaoticity} in the system as a mean damage size caused by a single node perturbation. If $\Omega>(<)0$ then mean damage size grows(shrinks) in time. Condition $\Omega=0$ will allow us to derive the relation for the critical line in the whole network.

Let us now divide the network into three parts: giant connected component (GC), finite clusters (FCs) and isolated points (IPs). The figure \ref{fig8} shows schematically how the single node perturbation evolves in time in these three parts of the network. In studied range of parameters the giant component behaves chaotically, i.e. the mean damage size is larger than initial perturbation and $\Omega_{GC}>0$. On the other hand, the small density of connections in finite tree-like clusters does not allow perturbation to spread out and $\Omega_{FCs}<0$. Because the state of isolated nodes does not change in time, then $\Omega_{IPs}=0$. Now, if one perturb randomly a set of nodes in the whole network, fraction $S$ of perturbations will be located in GC, fraction $(1-S)(1-P(k=0))$ will be located in FCs, and the rest of them, i.e. $(1-S)P(k=0)$ will perturb isolated nodes. Now one can write the condition for transition from the frozen to the chaotic state in the whole network:
\begin{equation}
S\Omega_{GC}+(1-S)(1-P(k=0))\Omega_{FCs}=0,
\end{equation}
where $\Omega$, $P(k)$ and $S$ depend on $\langle k\rangle$. This equation shows that the ordered behavior of small clusters can shield the chaotic behavior of the giant component. Only when {\it chaoticity} in GC is sufficiently developed, this shielding effect becomes neglected.

\begin{figure}[h]
\begin{center}
\includegraphics[width=8cm]{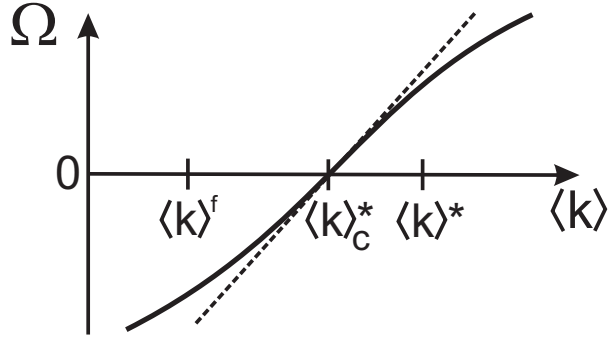}
\end{center}
\caption{{\it Chaoticity} $\Omega$ (solid line) in the viciity of the critical point $\langle k\rangle^*_c$. $\langle k\rangle^*$ and $\langle k\rangle^f$ are the average node degree in GC and in FCs respectively. Dashed line presents linear approximation of $\Omega$.}    \label{fig9}
\end{figure}

Now, expanding $\Omega$ into power series at $\langle k\rangle=\langle k\rangle_c$
\begin{equation}
\Omega=\Omega_0+\frac{\partial \Omega}{\partial \langle k\rangle}(\langle k\rangle-\langle k\rangle_c),
\end{equation}
where $\Omega_0=0$ in critical point (cf. Fig. \ref{fig9}), one gets the final equation for the critical line:
\begin{equation}
\label{final}
S(\langle k\rangle^*-\langle k\rangle^*_c)=(S-1)(1-P(k=0))(\langle k\rangle^f-\langle k\rangle^*_c),
\end{equation}
where $\langle k\rangle^f=\langle k\rangle u$ (cf. eq.(25) in \cite{bialas}).
The numerical solution of this implicit equation is presented in Fig. \ref{fig7} as the solid line.

\section{Conclusions}
This study was done to investigate the properties of undirected KBN model in the vicinity of
percolation threshold. We derived a mean field formula for the critical line characterizing KBN model in undirected and uncorrelated random graphs with arbitrary degree distributions. We have shown that the results of classical mean field theory differ from these obtained by numerical simulations. We have shown also that, to explain the discrepancies one has to take into account the effect of correlations between adjoining nodes in the giant connected component as well as the effect of shielding by finite size clusters. As one can see the problem is not easy even for undirected networks. As we have shown in figure 1 and 2 a directedness of the network introduces further complications into calculations. Nevertheless, we think that a similar approach can be derived even for that case. We hope that the presented work will encourage others to pursue these topics in the near future.

\section{Acknowledgments}
The work was funded in part by the European Commission Project CREEN FP6-2003-NEST-Path-012864
(P.F.), and by the Ministry of Education and Science in Poland under Grant
134/E-365/6.PR UE/DIE 239/2005-2007 (A.F.).

\end{document}